\documentclass[12pt]{article}
\usepackage{amsmath}
\usepackage{amsfonts}
\usepackage{amscd}
\usepackage{amsthm}
\usepackage{amssymb}
\usepackage{latexsym}
\usepackage{graphicx}
\usepackage{graphics}
\usepackage{subeqnarray}

\newcommand{\nc}{\newcommand*}

\usepackage[cp1251]{inputenc}
\inputencoding{cp1251}
\usepackage[T2B]{fontenc}
\usepackage[russian,english]{babel}
\nc{\beq}{\begin{equation}}
\nc{\eeq}{\end{equation}}
\nc{\lb}[1]{\label{#1}}
\nc{\nn}{\nonumber}
\nc{\ts}{\textstyle}
\nc{\ds}{\displaystyle}
\nc{\lar}{\leftarrow}
\nc{\rar}{\rightarrow}
\nc{\Rar}{\Rightarrow}
\nc{\half}{\frac{1}{2}}
\def\ot{\otimes}

\newcommand{\xp}{X^{+}}
\newcommand{\xm}{X^{-}}
\nc{\reff}[1]{(\ref{#1})} % Put parentheses around equation references
\begin{document}

%Title:
%\begin{center}
{\Large\bf  Deformed Richardson-Gaudin model} \\
%\end{center}

%Authors:
{\bf P. Kulish (1), A. Stolin (2), and H. Johannesson (3)}\\

% Affiliations

(1) Steklov Mathematical Institute, 27, Fontanka, 
191023 St. Petersburg, 
Russia
% and

(2) Department of Mathematics, Chalmers University of Technology and University\\ \indent \ \ \ \ \ of Gothenburg, SE 412 96 Gothenburg, Sweden

(3) Department of Physics, University of Gothenburg, SE 412 96 Gothenburg, Sweden \\

% "Alexander Stolin" <astolin@chalmers.se>:
%  lars.henrik.johannesson@gmail.com [lars.henrik.johannesson@gmail.com] f&#246;r Henrik Johannesson % % [henrik.johannesson@physics.gu.se]

\begin{quote}
{\bf Abstract.} The Richardson-Gaudin model describes strong pairing correlations of fermions confined to a finite chain. The integrability of the Hamiltonian allows for its eigenstates to be constructed algebraically. In this work we show that quantum group theory provides a possibility to deform the Hamiltonian while preserving integrability. More precisely, we use the so-called Jordanian r-matrix to deform the Hamiltonian of the Richardson-Gaudin model. In order to preserve its integrability, we need to insert a special nilpotent term into the auxiliary L-operator which generates integrals of motion of the system. Moreover, the quantum inverse scattering method enables us to construct the exact eigenstates of the deformed Hamiltonian. These states have a highly complex entanglement structure which requires further investigation. \\ 
\end{quote}

PACS numbers: 02.30.Ik, 03.65.Fd, 21.10.Re, 74.20.Rp
\newpage

%\section{Introduction}
The Richardson-Gaudin model \cite{1,4} is an integrable spin-1/2 periodic chain with Hamiltonian
\begin{equation}\label{1}
H=\sum_{j=1}^N \epsilon_j S^z_j + g\sum_{j,\,k=1}^N S^-_j S^+_k,
\end{equation}
where $g$ is a coupling constant and
$S^{\pm}_l = (S^x_l \pm iS^y_l)$, with $S^{\alpha}_l$ $N$ copies of the Lie algebra $su(2)$ generators
\begin{equation}\label{2}
[S^{\alpha}_l,S^{\beta}_{l'}]=i\varepsilon^{\alpha\,\beta\,\gamma}S^{\gamma}\delta_{l\,l'}, \ \ \ \  \alpha, \beta, \gamma = x,y,z.
\end{equation}
As shown by Cambiaggio {\em et al.} \cite{Cambiaggio}, by introducing fermion operators  $c^{\dagger}_{lm}$ and $c_{lm}$
related to the sl(2) generators by
\begin{equation}
S^z_l = \frac{1}{2}\sum_m c^{\dagger}_{lm} c_{lm} - \frac{1}{2}, \ \ \  S^+_l = \frac{1}{2}\sum_m c^{\dagger}_{lm} c^{\dagger}_{l\bar{m}} = (S^-_l)^{\dagger},
\end{equation}
the Richardson-Gaudin model in Eq. (\ref{1}) gets mapped onto the pairing model Hamiltonian
\begin{equation} \label{pairing}
H_P = \sum_l \epsilon_l \hat{n}_l + \frac{g}{2}\sum_{l, l'} A^{\dagger}_l A_{l'}.
\end{equation}
Here $c^{\dagger}_{lm}$ ($c_{lm}$) creates (annihilates) a fermion in the state $\mid\! l m \rangle$ (with $\mid\! l \bar{m} \rangle$ the time reversed state of
$\mid\! l m \rangle$), and
$n_l = \sum_m c^{\dagger}_{lm}c_{lm}$ and $A^{\dagger}_l = (A_l)^{\dagger} = \sum_m c^{\dagger}_{lm} c^{\dagger}_{l\bar{m}}$ are the corresponding
number- and pair-creation operators. The pairing strengths $g_{l l'}$ are here approximated by a single constant $g$, with $\epsilon_l$ the single-particle level
corresponding to the $m$-fold degenerate states $\mid \!l m\rangle$.

As is well-known, the pairing model in Eq. (\ref{pairing}) is central in the theory of superconductivity. Richardson's exact solution of the model \cite{1}, exploiting
its integrability, has been important for applications in mesoscopic and nuclear physics where the small number of fermions prohibits the use of conventional BCS theory \cite{RMP}. Moreover, its (pseudo)spin representation in the guise of
the Richardson-Gaudin model, Eq. (\ref{1}), provides a striking link between quantum magnetism and pairing phenomena, two central concepts in the physics of quantum matter.

The eigenstates of the Richardson-Gaudin Hamiltonian, Eq. (\ref{1}), can be constructed algebraically using the quantum inverse
scattering method (QISM) \cite{2,3}. The main objects of this method are the classical $r$-matrices
\begin{equation}\label{3}
r(\lambda,\,\mu)=\frac{4}{\lambda-\mu}\sum_{\alpha} S^{\alpha}\ot S^{\alpha}\rule[-10pt]{0.2pt}{20pt}_{\,s=\frac{1}{2}}\simeq \frac{1}{\lambda - \mu}
\begin{pmatrix}
1&0&0&0\\
0&-1&2&0\\
0&2&-1&0\\
0&0&0&1
\end{pmatrix}.
\end{equation}
and the $L$-matrix of the loop algebra $\mathcal{L}(sl(2))$ generators $h(\lambda),\,X^+(\lambda),\,X^-(\lambda)$
\begin{equation}\label{4}
L(\lambda)=
\begin{pmatrix}
h(\lambda)&2X^-(\lambda)\\
2X^+(\lambda)&-h(\lambda)
\end{pmatrix}.
\end{equation}

The commutation relations (CR) of loop algebra generators are given in compact matrix form
\begin{equation}\label{5}
[L_1(\lambda),L_2(\mu)]=-[r_{1\,2}(\lambda,\mu),L_1(\lambda)+L_2(\mu)],
\end{equation}
where $L_1(\lambda)=L(\lambda)\ot\mathbb{I}$,  $L_2(\mu)=\mathbb{I}\ot L(\mu)$ and $r(\lambda,\,\mu)$ is the $4\times 4$ $c$-number matrix in Eq. \reff{3}. A consequence of this form is the commutativity of transfer matrices,
\begin{equation}\label{6}
t(\lambda)=\half\text{tr}_0(L^2(\lambda))\in \mathcal{L}(sl(2)),\qquad
[t(\lambda),t(\mu)]=0.
\end{equation}

The corresponding mutually commuting operators extracted from the decomposition of $t(\lambda)$ define a Gaudin model \cite{4,5}. However, to get the Richardson Hamiltonian a mild change of the $L$-operator is necessary,
$$
L(\lambda)\rightarrow L(\lambda;c):= c\, h_0+L(\lambda), %\equiv Y(\lambda),
$$
where $h_0=\sigma_0^z$ in auxiliary space $\mathbb{C}_0^2$ of spin 1/2.
This transformation does not change the CR of matrix elements of this matrix $L(\lambda;c)$
due to the symmetry of the $r$-matrix \reff{3}:
\begin{equation}\label{7}
[Y\ot\mathbb{I}+\mathbb{I}\ot Y, r(\lambda,\,\mu)]=0,\quad Y\in sl(2).
\end{equation}
The resulting transfer matrix obtains some extra terms
\begin{equation} \label{8}
t(\lambda;c)=\half\text{tr}_0\left(L(\lambda;c)\right)^2=
c^2\mathbf{1}+c\, h(\lambda)+h^2(\lambda)
+2\left( X^+(\lambda)X^-(\lambda)+X^-(\lambda)X^+(\lambda) \right).
%X^+(\lambda).
\end{equation}

Let us consider a spin-$1/2$ representation on the auxiliary space
$V_0\simeq\mathbb{C}^2$ and spin $\ell_{k}$ representations on quantum spaces
$V_k\simeq\mathbb{C}^{\ell_{k}+1}$ with extra parameters $\epsilon_k$ corresponding to site
$k=1,2,\ldots,N$. The whole space of quantum states  is $\mathcal{H}=\ds{\otimes_1^N}V_k$
and the highest weight vector (highest spin, ''ferromagnetic state'') $\mid\!\Omega_+\rangle$ satisfies
\begin{equation}\label{93}
X^+(\lambda)\mid\!\Omega_+\rangle=0,\qquad h(\lambda)\mid\!\Omega_+\rangle=\rho(\lambda)\mid\!\Omega_+\rangle,
\end{equation}
where
$$
\rho(\lambda)=\sum_{k=1}^N\frac{l_{k}}{\lambda-\epsilon_k}.
$$
It is useful to introduce the notation $Y_{gl}$ for global operators of the $sl(2)$-representation:
$$
Y_{gl}:=\sum_{k=1}^N Y_k.
$$
To find the eigenvectors and spectrum of $t(\lambda)$ on
$\mathcal{H}$ one requires that vectors of the form
\begin{equation}\label{10}
|\mu_1,\ldots , \mu_M\rangle=\prod_{j=1}^M X^-(\mu_j)\mid\!\Omega_+\rangle
\end{equation}
are eigenvectors of $t(\lambda)$,
\begin{equation}\label{11}
t(\lambda)|\{\mu_j\}_{j=1}^M\rangle=\Lambda(\lambda;\{\mu_j\}_{j=1}^M)|\{\mu_j\}_{j=1}^M\rangle,
\end{equation}
provided that the parameters $\mu_j$ satisfy the Bethe equations:
\begin{equation}\label{12}
2c+\sum_{k=1}^{N}\frac{\ell_{k}}{\mu_{i}-\epsilon_k} - \sum_{j\neq i}^{M}\frac{2}{\mu_{i}-\mu_{j}} = 0,\quad i=1,\dots,M.
\end{equation}
The realization of the loop algebra generators on the space $\mathcal{H}$ takes the form
\begin{equation}\label{13}
h(\lambda) = \sum_{k=1}^{N} \frac{h_{k}}{\lambda - \epsilon_k} ,\quad
\xm(\lambda) = \sum_{k=1}^{N} \frac{\xm_{k}}{\lambda - \epsilon_k},\quad
\xp(\lambda) = \sum_{k=1}^{N} \frac{\xp_{k}}{\lambda - \epsilon_k}.
\end{equation}

The coupling constant $g$ of \reff{1} is connected with the parameter $c=1/g$, while the Hamiltonian \reff{1} is obtained as the operator coefficient of the term $1/\lambda^2$ in the expansion of $t(\lambda;c)$ at $\lambda\rightarrow\infty$.

Quantum group theory gives the possibility to deform a Hamiltonian preserving integrability
\cite{6,7}. Specifically, we can use the so-called Jordanian $r$-matrix to quantum deform the Hamiltonian of Richardson-Gaudin model \reff{1}.
We add to the $sl(2)$ symmetric $r$-matrix \reff{3} the Jordanian part
\begin{equation}\label{14}
r^{J}(\lambda,\,\mu)=\frac{C_2^{\ot}}{\lambda-\mu}+\xi\left(h\ot\xp-\xp\ot h\right),
\end{equation}
with Casimir element $C_2^{\ot}$ in the tensor product of two copies of $sl(2)$,
$$
C_2^{\ot}=h\ot h+2\left(\xp\ot\xm+\xm\ot\xp\right).
$$

After the Jordanian twist the r-matrix (14) is commuting with the
generator $X_0^{+}$ only,
\begin{equation}\label{15}
[X_0^{+} \otimes \mathbb{I} + \mathbb{I} \otimes X_0^{+} , r^{(J)}(\lambda, \mu)] = 0.
\end{equation}
Hence, one can add the term $c X_0^{+} + L(\lambda, \xi) $ to the $L$-operator.
This yields the twisted transfer-matrix
$t^{(J)}(\lambda) = {\frac12 }\mbox{tr}_0 (c X_0^{+} + L(\lambda, \xi) )^2$,
\begin{equation}\label{16}
t^{(J)}(\lambda) = c X^{+}(\lambda) + h(\lambda)^2 - 2h'(\lambda) + 2(2 X^{-}(\lambda) +\xi) X^{+}(\lambda).
\end{equation}
The corresponding commutation relations between the generators of the twisted
loop algebra are explicitly given by
\begin{align}
\left[h(\lambda),h(\mu)\right] &= 2\xi \left(\xp(\lambda) - \xp(\mu)\right)\notag\\
\left[\xm(\lambda),\xm(\mu)\right] &= -\xi \left(\xm(\lambda) - \xm(\mu) \right),\notag \\
\left[\xp(\lambda),\xm(\mu)\right] &= -\frac{h(\lambda) - h(\mu)}{\lambda - \mu} + \xi \xp(\lambda),\notag\\
\left[\xp(\lambda),\xp(\mu)\right] &= 0,
\label{17}
\\
\left[h(\lambda),\xm(\mu)\right] &= 2\frac{\xm(\lambda) - \xm(\mu)}{\lambda - \mu} + \xi h(\mu),\notag\\
\left[h(\lambda),\xp(\mu)\right] &= -2\frac{\xp(\lambda) - \xp(\mu)}{\lambda - \mu}.\notag
\end{align}

The realization of the Jordanian twisted loop algebra ${\cal L}_J(sl(2))$ with CR \reff{17} is given similar to \reff{13}
with extra terms proportional to the deformation parameter $\xi$,

\begin{gather}
\label{18}%{semiclassicalgenerators}%%%%%%%
h(\lambda) = \sum_{k=1}^{N} \left(\frac{h_{k}}{\lambda - \epsilon_k} + \xi \xp_{k}\right),\;
\xm(\lambda) = \sum_{k=1}^{N} \left(\frac{\xm_{k}}{\lambda - \epsilon_k} - \frac{\xi}{2} h_{k}\right),\;
\xp(\lambda) = \sum_{k=1}^{N} \frac{\xp_{k}}{\lambda - \epsilon_k}.
\end{gather}

To construct eigenstates for the twisted model one has to use operators of the form \cite{7,8}
\begin{equation}\label{19}
    B_M (\mu_1,...,\mu_M) = X^-(\mu_1) (X^-(\mu_2) + \xi) ...(X^-(\mu_M) + \xi (M-1)),
\end{equation}
and acting by these operators on the ferromagnetic state $\mid\!\Omega_+\rangle$.

The deformed Richardson-Gaudin model Hamiltonian can now be extracted from the transfer-matrix $t^{(J)}(\lambda)$ as
the operator coefficient in its expansion $\lambda \to \infty $.

According to Eqs. \reff{6} and \reff{16} one can also extract quantum integrals of motion $J_k$ using the realization \reff{18},
reading off the expressions for $J_k$ from the expansion
\begin{equation}\label{20}
t^{(J)}(\lambda)=J_0+\frac{1}{\lambda}J_1+\frac{1}{\lambda^2}J_2+\ldots.
\end{equation}
The corresponding quantum deformed Hamiltonian reads
\begin{equation}
\quad H\simeq J_2=c\sum_{j=1}^N\epsilon_jX_j^+
+2\xi\left\{\left(\sum_{j=1}^N\epsilon_jh_j\right)X^+_{gl}-h_{gl}\sum_{j=1}^N\epsilon_jX_j^+\right\} +
\left(h_{gl}^{\,\,\,2}+2h_{gl}+4X^-_{gl}X^+_{gl}\right).
\end{equation}
It is instructive to write down a simplified case without the Jordanian twist: $\xi=0$. One thus obtains
\begin{equation}\label{21}
J_0=0,\quad  J_1=X^+_{gl},\quad  J_2\simeq\sum_{k=1}^N \epsilon_k X_k^+
+\frac{g}{2}\left(h^{\,\,\,2}_{gl}+2h_{gl}+4X^-_{gl}X^+_{gl}\right).
\end{equation}

The case $\xi =0$ can also be obtained by taking off from the inhomogeneous  $XXX$ spin chain. The model can be described by a $2\times 2$ monodromy 
matrix \cite{2}
\begin{equation*}
 T(\lambda)=\begin{pmatrix} A(\lambda)&B(\lambda)\\ C(\lambda)& D(\lambda) \end{pmatrix},
\end{equation*}
satisfying the quadratic relations
\begin{equation}\label{22}
R(\lambda,\mu)T(\lambda)\otimes T(\mu)=
\left(I\otimes T(\mu)\right)\left(T(\lambda)\otimes I\right)R(\lambda,\mu).
\end{equation}
If we multiply $T(\lambda)$ by a constant $2\times 2$ matrix ${\cal M}$ the resulting matrix $\widetilde{T}(\lambda)={\cal M} \cdot T(\lambda)$ will satisfy the same relation \reff{22}. Choosing a triangular 
matrix 
\begin{equation*}
    {\cal M}(\epsilon)=\begin{pmatrix} 1&\varepsilon \\ 0& 1 \end{pmatrix},
\end{equation*}
the entries of monodromy matrices become simply related:
\begin{equation*}
 \widetilde{A}=A+\varepsilon C,\quad \widetilde{B}=B+\varepsilon D,\quad 
 \widetilde{C}=C,\quad \widetilde{D}=D.
\end{equation*}
This choice of ${\cal M}(\epsilon)$ (of the same type as considered in Ref. \cite{11}) permits us to use the same reference state  $\mid\!\Omega_+\rangle\in\mathcal{H}$ \reff{93} and
$\widetilde{B}$ as a creation operator of the algebraic Bethe ansatz \cite{2}.

Bethe states are given by the same action of product operators  
$ \widetilde{B}(\mu_{j})=B(\mu_{j})+\varepsilon D(\mu_{j})$
although operators $B(\mu_{j})$ do not commute with $ D(\mu_{j})$:
\begin{equation}\label{DB}
D(\lambda)B(\mu)=\alpha(\lambda, \mu)B(\mu)D(\lambda)+\beta(\lambda, \mu)B(\lambda)D(\mu),
\end{equation}
where
\begin{equation*}
\alpha(\lambda, \mu)=\frac{\lambda-\mu+\eta}{\lambda-\mu},\qquad
\beta(\lambda, \mu)=-\frac{\eta}{\lambda-\mu}.
\end{equation*}
For a 3-magnon state one gets due to B-D ordering in Eq. (\ref{DB}):
\begin{multline*}
\prod_{j=1}^3 \widetilde{B}(\mu_{j})=\prod_{j=1}^3 B(\mu_{j}) +\varepsilon
\sum_{s=1}^3  \alpha(\mu_k, \mu_s) \alpha(\mu_s, \mu_l)B(\mu_{k})B(\mu_{l})D(\mu_{s})+\qquad\\
\varepsilon^2\sum_{s=1}^3  \alpha(\mu_k, \mu_s) \alpha(\mu_l, \mu_s)B(\mu_{s})D(\mu_{k}))D(\mu_{l})+
\varepsilon^3\prod_{j=1}^3D(\mu_{j}).\qquad
\end{multline*}
Similar formulas are valid for $M$-magnon states. 
Hence, acting on the ferromagnet state  $\mid\!\Omega_+\rangle$ we obtain filtration of states with eigenvalues of 
$S^z: \frac{N}{2}, \frac{N}{2} -1,\frac{N}{2}-2,\frac{N}{2}-3$.

More complicated deformations of the Richardson-Gaudin model
can be obtained using r-matrices related to the higher rank Lie algebras \cite{9}.
The structure of the eigenstates of the transfer matrix and their entanglement properties
\cite{10} are under investigation.
\bigskip

{\bf Acknowledgments}
We would like to thank E. Damaskinsky for useful discussions. This work was supported by RFBR grants 11-01-00570-a, 12-01-00207-a, and 13-01-12405-ofi-M2 (P.K.), and by STINT grant IG2011-2028 (A.S and H.J.). \\

\end{document}